\documentclass{article}

\usepackage[headinclude,footinclude]{typearea}

\typearea{10}
\usepackage{amssymb}       % Use the fancy AmsFonts 

\usepackage{graphicx}      % PostScript graphics, extended syntax
\usepackage{caption}
\usepackage{subcaption}

\begin{document}

\begin{center} 
{\Large{\bf Present Status of the EMC effect 
\footnote{Lecture notes prepared for the
Proceedings of the International School of Subnuclear Physics, Erice, \\
24 June -- 3 July 2013}
\\}}
\vspace{1cm}
{\large {\bf Klaus Rith}}\\

\vspace{10 mm}

             Friedrich-Alexander-Universit\"at Erlangen-N\"urnberg\\
             Physics Department\\
             Erwin-Rommel-Str. 1\\
             D-91058 Erlangen, Germany \\
             E-mail: klaus.rith@physik.uni-erlangen.de

\end{center}

\begin{abstract}
The present status of the EMC effect, the modification of the per nucleon cross section
in deep-inelastic lepton nucleus scattering by the nuclear environment, is reviewed.
\end{abstract}

\section{Introduction}
The 30$^{\rm th}$ anniversary of the publication by the European Muon Collaboration (EMC)
\cite{Aubert83} on the modification of the per-nucleon cross section of nucleons bound in
nuclei, named 
{\it EMC effect}, is a good opportunity to review the status of the effect and to summarize the 
available experimental and theoretical information. The EMC effect can be interpreted as a 
modification of quark and gluon distributions in bound nucleons by the nuclear environment.
The dependences of these nuclear modifications on kinematics and various nuclear properties like 
mass, density or radius are meanwhile rather well known, nevertheless the origin of the effect 
is still not fully understood. Recent data, discussed in section \ref{xlargerone}, provide new 
important input. 

In my opinion further
measurements of the effect in electron-nucleus, neutrino-nucleus and proton-nucleus scattering
are needed to determine nuclear quark and gluon distributions (nPDFs) down to very low parton 
momentum fractions $x$ and to constrain the initial state for the ${\rm AA}$ program at RHIC and 
LHC, or for the correct interpretation of, e.g., (future) $\nu{\rm A}$ experiments.

\section{Deep-inelastic scattering, structure functions and quark distributions}
\label{kin}
Nuclear modifications of parton distributions have been studied mainly by measurements
of cross sections in deep-inelastic lepton-nucleus scattering. For electromagnetic interactions 
of charged leptons with nuclear targets and in the approximation of one-photon exchange the 
cross section reads 
\begin{equation}
\frac {d^2\sigma}{d Q^2 d x} 
= \frac{4\pi\alpha^2 }{Q^4} \frac{{\rm F}_2(x,Q^2)}{x}
 \left[1 - y - \frac{Q^2}{4{\rm E}^2}
 + \frac{y^2 + Q^2/{\rm E}^2}{2\left [1 + {\rm R}(x,Q^2)\right ]} \right ] \;. 
\label{gl:DIScross}
\end{equation}
Here $-Q^2$ represents the squared four-momentum of the virtual photon that mediates the 
interaction with coupling strength $\alpha$ and $x = Q^2/2{\rm M}\nu$ can be interpreted as the 
fraction of the longitudinal nucleon momentum carried by the struck quark, in a frame where the 
nucleon moves with infinite momentum in the direction opposite to that of the virtual photon. 
The variable $y$ denotes, in the target rest frame , the virtual-photon energy $\nu$ with 
respect to the lepton-beam energy $\rm E$.

At leading order in QCD the structure function ${\rm F}_2$ is defined as  
the sum of the momentum distributions 
${\rm q}(x,Q^2)$ and $\bar{\rm q}(x,Q^2)$
of quarks and anti-quarks of flavor 
$\rm q = u, d, s, ...$ weighted by $x$ and $z_{\rm q}^2$, where $z_{\rm q}$ is the quark charge
(in units of the elementary charge $|e|$):
\begin{equation} 
{\rm F}_2(x,Q^2) = \sum_{\rm q=u,d,s..}xz_{\rm q}^2 \left[ {\rm q} (x,Q^2) + \bar{\rm q} (x,Q^2) 
\right] \;.
\label{gl:f2quarks} 
\end{equation}

\noindent The quantity 
\begin{equation}
{\rm R} = \frac{\sigma_{\rm L}}{\sigma_{\rm T}} 
=\frac{{\rm F}_2}{2x{\rm F}_1}\left[1 + \frac{4{\rm M}^2x^2}{Q^2} \right] - 1  = 
\frac{{\rm F}_{\rm L}}{2x{\rm F}_1}
\label{gl:R} 
\end{equation}
 is the ratio of the longitudinal to transverse virtual-photon cross sections. In the  
quark-parton model, $\rm R = 0$ for the interaction of the 
virtual photon with a point-like spin-1/2 particle. Quark transverse momenta, quark masses and 
gluon radiation cause $\rm R$ to deviate from zero. If $\rm R$ is independent of the nuclear 
mass number $\rm A$ (see the discussion in section \ref{RAminusRD}), then the ratio of cross 
sections for two 
different nuclei is equal to the ratio of their structure functions ${\rm F}_2$. 

Subsequently, we will always discuss the ratio of structure functions (cross sections) per 
nucleon for a nucleus
with mass number $\rm A$ (i.e., $\rm A$ nucleons) and the deuteron $\rm D$. The latter is, to a 
good approximation,  equal to
the proton-neutron averaged structure function 
${\rm F}_2^{\rm D}\approx ({\rm F}_2^{\rm p}+{\rm F}_2^{\rm n})/2$.
The $x$ dependence of the structure functions ${\rm F}_2^{\rm p}$ and ${\rm F}_2^{\rm n}$ is 
different (for 
free nucleons they are approximately related by 
${\rm F}_2^{\rm n}/{\rm F}_2^{\rm p} \approx 1 - 0.8 x$). 
Results for the nuclear structure function ${\rm F}_2^{\rm A}$ (cross section 
$\sigma^{\rm A}$) for nuclei with $\rm Z$ protons and $\rm N$ neutrons will always be corrected 
for neutron excess by
\begin{equation} 
{\rm F}_2^{\rm A} =( \frac{{\rm F}_2^{\rm p}+{\rm F}_2^{\rm n}}{2})^{\rm A} \cdot 
\left[1 - \frac{{\rm N}-{\rm Z}}{{\rm N}+{\rm Z}}\cdot 
\frac{1 - {\rm F}_2^{\rm n}/{\rm F}_2^{\rm p}}{1 + {\rm F}_2^{\rm n}/{\rm F}_2^{\rm p}}
\right],
\label{gl:f2A}
\end{equation}
where it is assumed that proton and neutron structure functions are modified equally by the 
nuclear environment.
Thus, ${\rm F}_2^{\rm A}$ is the structure function per nucleon for a hypothetical isoscalar 
nucleus with an equal number ($\frac{\rm A}{2}$) of protons and neutrons.

\section{The discovery}

\begin{figure}[h!] %[p]
\centering
\includegraphics[scale=0.07]{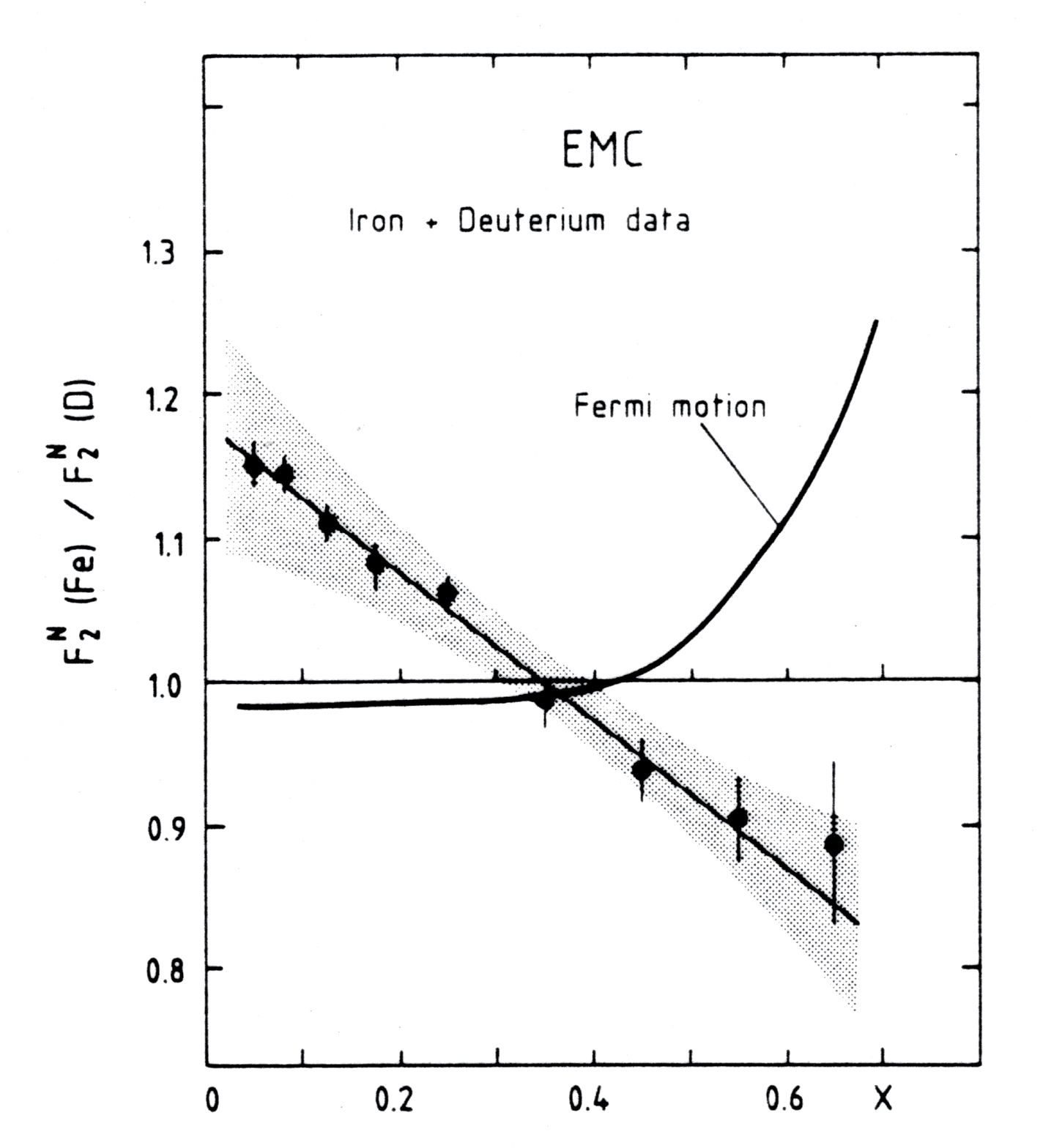}
\includegraphics[scale=0.18]{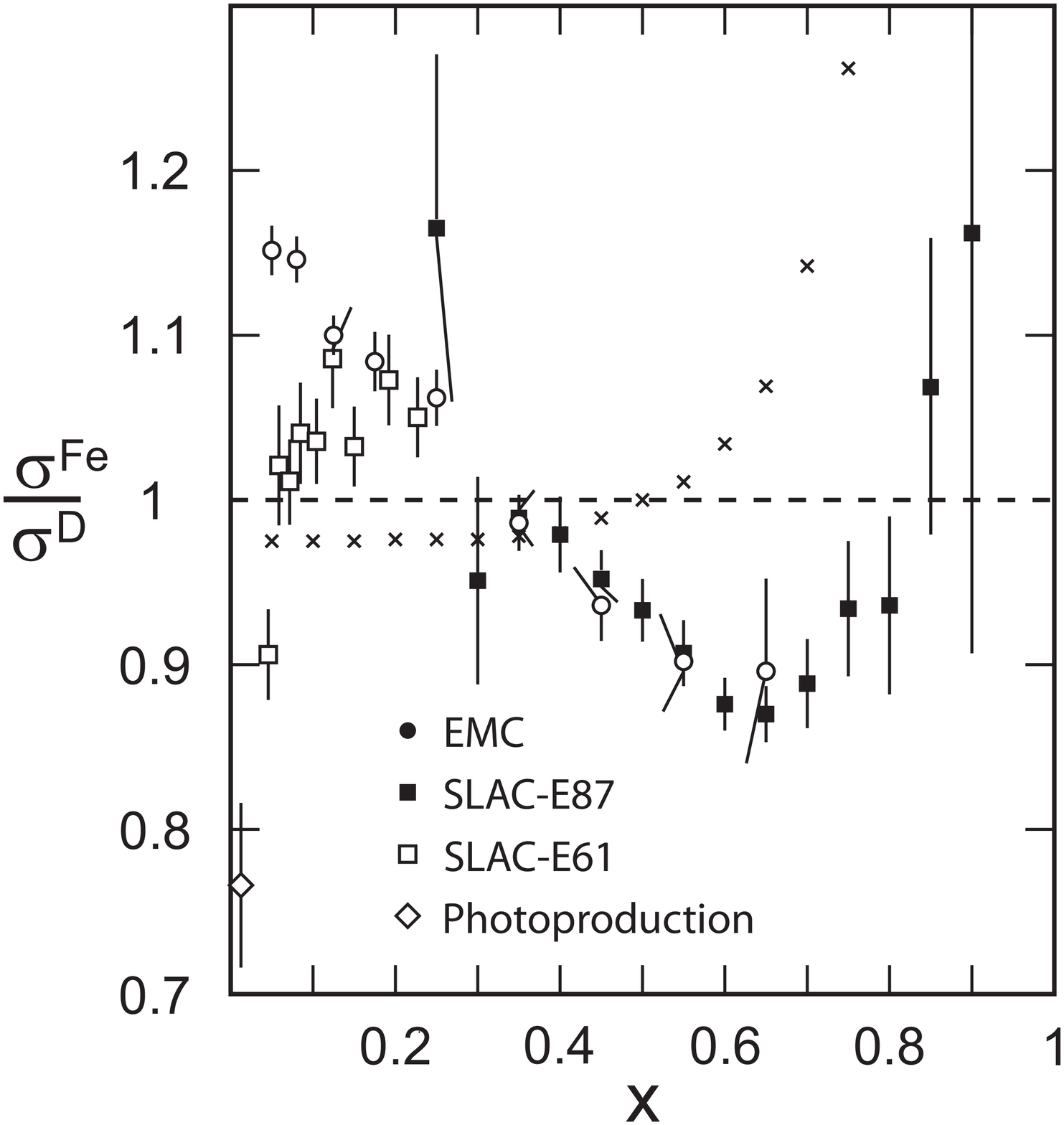}
\caption{%
   The ratio ${\rm F}_2^{\rm Fe}/{\rm F}_2^{\rm D}$ ($\sigma^{\rm Fe}/\sigma^{\rm D}$) as a 
function of $x$. Left panel: the original EMC result~\cite{Aubert83}; right panel: The SLAC 
result \cite{Bodek83a} together with low-$Q^2$ SLAC data for Cu/D~\cite{Stein75} and the 
photoproduction result
at $Q^2 = 0$ GeV$^2$~\cite{Caldwell79}. Also shown is one expectation for the effect of Fermi 
motion on ${\rm F}_2^{\rm Fe}$  in absence of other nuclear effects~\cite{Bodek81}.
}
\label{ab:history}
\end{figure}

The historical result of the EMC effect~\cite{Aubert83} (updated results were published in
\cite{Aubert87}) is presented in the left panel of
Fig. \ref{ab:history}. It shows the ratio of the structure function ${\rm F}_2$ per nucleon for 
iron and deuterium, both uncorrected for Fermi motion, as a function of $x$. The shaded area indicates the range for the errors on the slope of a linear fit to the data, the point-to-point systematic uncertainties are somewhat larger. In addition there is an overall uncertainty of $\pm 7 \%$.

The ratio is seen to be different from unity. It falls from 
$\sim1.15$ at $x = 0.05$ to a value of $\sim0.89$ at $x = 0.65$ and doesn't follow the 
expectations from Fermi-motion calculations. This result demonstrated 
for the first time that the structure function ${\rm F}_2$ is
modified when nucleons are embedded in a nucleus. At the time when these data were presented  
the effect appeared quite astounding. Especially members of the high-energy physics community 
could 
hardly believe that at momentum transfers several orders of magnitude larger than typical 
nuclear binding energies quark distributions should be affected by the nuclear environment. 
However, `quarks in nuclei' and phenomena that could possibly modify the quark distributions in 
bound 
nucleons were already discussed by the nuclear-physics community for some time before this 
discovery.
(For a first review of such ideas see Ref. \cite{Rith83}).
The EMC result was quickly confirmed by the SLAC-MIT-Rochester group that recovered and 
reanalyzed the data stemming from the aluminum and steel cell walls of the liquid hydrogen 
and deuterium targets
from the experiments E49B and E87. The result for $\sigma^{\rm Fe}/\sigma^{\rm D}$ 
\cite{Bodek83a} is shown in the right panel of  Fig. \ref{ab:history} together with earlier data 
from SLAC-E61 at lower $Q^2$ for Cu/D
in the range $0.04 < x < 0.25$ \cite{Stein75} and a data point at $Q^2 = 0$ GeV$^2$ from an 
experiment that had investigated shadowing in photoproduction~\cite{Caldwell79}. (Similar data 
were published for aluminum~\cite{Bodek83b}). At high $x$,
these data were in good agreement with the EMC data, but the rather large value of
${\rm F}_2^{\rm Fe}/{\rm F}_2^{\rm D}$ for the two
low-$x$ EMC points was in disagreement with the low-$Q^2$ SLAC data which indicated that for $x$ 
below $\sim0.15$ the ratio decreases again with decreasing $x$. (Indeed, it was found out 
later that the low-$x$ EMC points suffered from correlated tracking 
inefficiencies affecting the deuterium but not the iron data).

This exciting result caused enormous activities in both experiment and theory, resulting at 
present in about 1000 citations of the EMC publication~\cite{Aubert83}. The experimental data 
(with the exception of the recent JLAB data presented in Sect.~\ref{xlargerone}) and many of 
the theory 
papers have been discussed in great detail some time ago in the excellent review by P. 
Norton~\cite{Norton03}.
Therefore, I will only summarize some of the key results and the main 
theoretical ideas for the interpretation of the effect.

\section{The experimental data}
After the discovery, nuclear effects have been studied experimentally in charged lepton-nucleus 
scattering
by the muon experiments BCDMS~\cite{bcdms}, EMC-NA38~\cite{emc-na38}, EMC~\cite{emc} and NMC~
\cite{Amaudruz91,Amaudruz92a,Amaudruz92b,Amaudruz92c,Arneodo95,
Arneodo96a,Arneodo96b} at CERN and E665~\cite{e665,Adams95} at FNAL, in electron scattering at 
SLAC~\cite{Arnold84,Gomez94,Dasu88,Dasu94,Frankfurt93}, DESY~\cite{Airapetian03} and JLAB~\cite
{Egyan03,Egyan06,Seely09,Fomin12}, in neutrino-nucleus scattering~\cite{cdhs,neutrinos-bc}
and in the Drell-Yan process~\cite{Alde90,Vasiliev99}. I will only discuss the most important 
results.

\subsection{Data from SLAC-E139}
\begin{figure}[h!] %[p]
\centering
\includegraphics[scale=0.22]{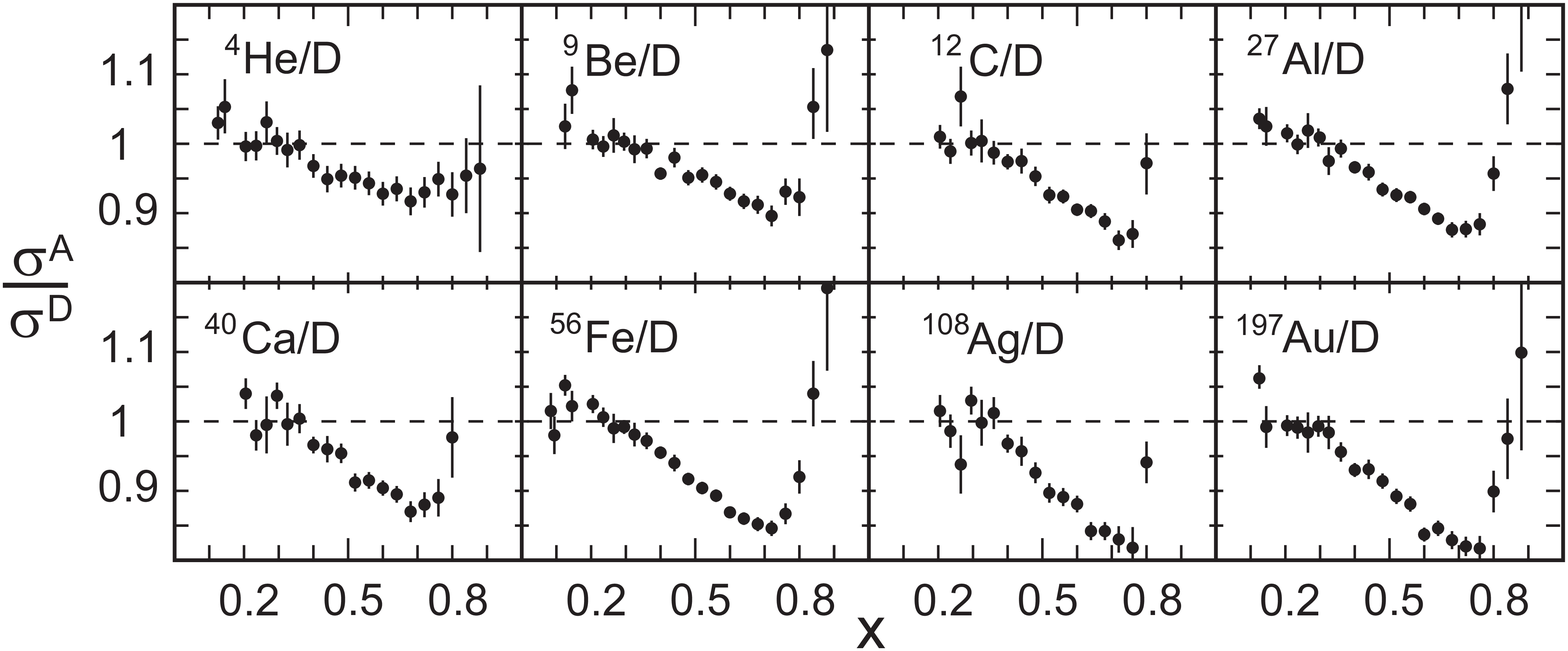}
\caption{%
   The ratio $\sigma^{\rm A}/\sigma^{\rm D}$ as a function of $x$ for  various nuclei measured by SLAC-E139 
\cite{Arnold84,Gomez94}.
} 
\label{ab:SLACE139}
\end{figure}

At large $x$ the most precise data are those from the SLAC experiment E139~\cite{Arnold84}
that measured the cross section ratio $\sigma^{\rm A}/\sigma^{\rm D}$ for 8 nuclei ranging from
$^4$He to $^{197}$Au. The results of an updated analysis with an improved treatment of radiative 
corrections~\cite{Gomez94} are shown in Fig. \ref{ab:SLACE139}.

For all nuclei one observes, in the region $0.3 < x < 0.8$, a reduction of the per-nucleon cross 
section  $\sigma^{\rm A}$ compared to the `free nucleon' one, $\sigma^{\rm D}$. The $x$ dependence
of this reduction has a very characteristic universal shape with a minimum near 
$x \approx 0.7$. The effect is already present for helium, its magnitude increases with the 
atomic mass number $\rm A$. The $\rm A$ dependence will be discussed in more detail in section
\ref{nuclprop}.

\subsection{The universal $x$ dependence}
\begin{figure}[h!] %[p]
\centering
\includegraphics[scale=0.22]{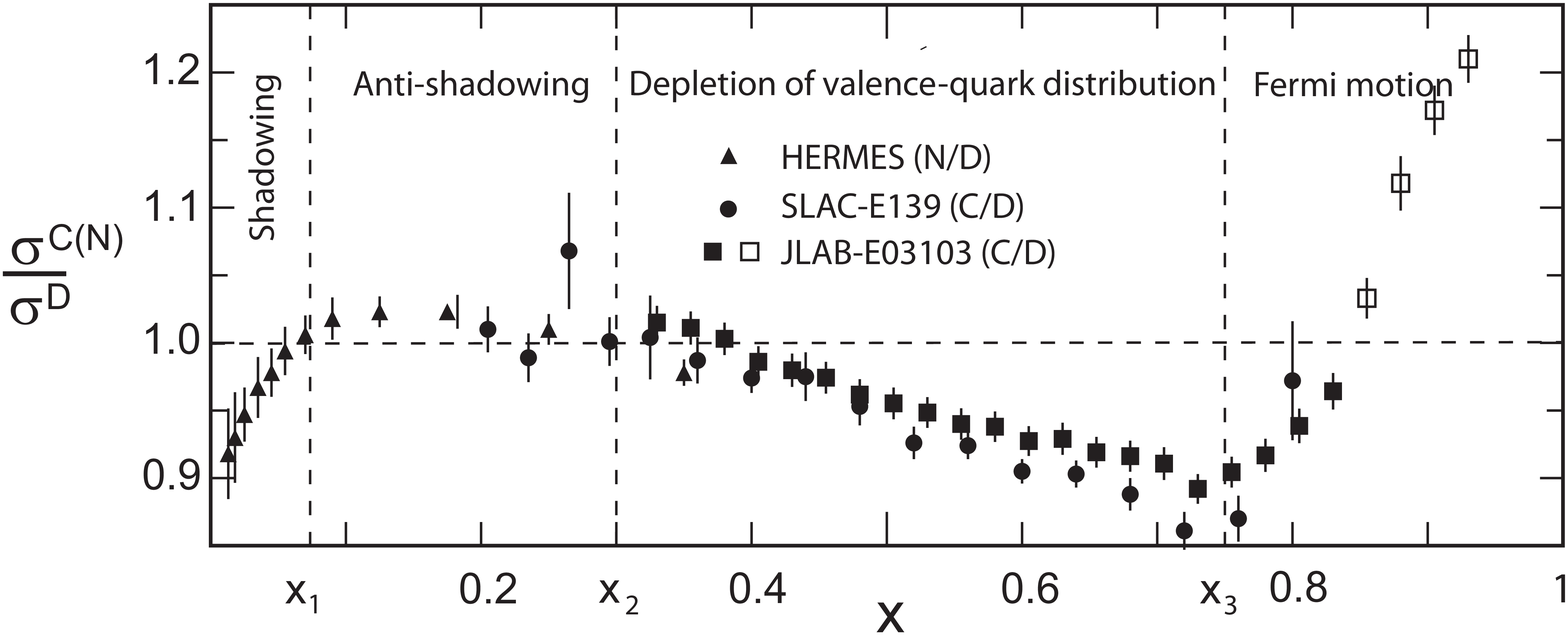}
\caption{%
   The ratio $\sigma^{\rm C(N)}/\sigma^{\rm D}$ as a function of $x$ from 
HERMES~\cite{Airapetian03}, SLAC-E139~\cite{Gomez94}, and JLAB-E03103~\cite{Seely09}. Open 
squares denote ${\rm W}^2$ below 2
GeV$^2$, where $\rm W$ is the invariant mass of the photon-nucleon system.
} 
\label{ab:globalx}
\end{figure}

The E139 data provide precise information for $x > 0.2$. The region of lower $x$ is covered by 
the data 
from SLAC-E61\cite{Stein75}, from the  HERMES experiment~\cite{Airapetian03}, 
where the 27.6 GeV electron beam of HERA was scattered from internal gas targets of various 
nuclear species, and the muon experiments. As an example, the $x$ dependence of the  cross 
section ratio $\sigma^{\rm C(N)}/\sigma^{\rm D}$ measured in electron scattering by E139~\cite
{Gomez94} and HERMES~\cite{Airapetian03} is presented in Fig. \ref{ab:globalx}. 
Also shown are data from JLAB-E03103~\cite{Seely09} taken with a beam energy of 5.8 GeV.

\noindent
This figure nicely summarizes the universal $x$ dependence of the nuclear effects. It can be 
subdivided
into four $x$ regions (plus a fifth one at $x > 1$ which will be discussed separately in section 
\ref{xlargerone}): 
\begin{itemize}
\item the 'shadowing' region ($0 < x < x_1\backsimeq 0.06$), where the structure function ratio 
is smaller than unity and decreases with decreasing $x$ down to the value measured in 
photoproduction. Here, the dominant contribution to the cross section is due to sea quarks. 
The essential longitudinal distances $\Delta{z}$ probed in the deep-inelastic interaction (see 
section \ref{shad}) are $\Delta{z} > 3$ fm, much bigger than the size of a nucleon;
\item the 'anti-shadowing' region ($x_1 < x < x_2\backsimeq 0.3, ~3~{\rm fm} > \Delta{z}> 
0.7~{\rm fm}$), where the ratio shows a small increase of a few percent over unity;
\item the region ($x_2 < x < x_3\backsimeq 0.8, ~\Delta{z} < 0.7~{\rm fm}$), where the ratio is 
smaller than unity with a minimum near $x \approx 0.7$. Here, the 
sea-quark distribution is essentially negligible and the ratio reflects the behavior of the 
valence-quark distributions; 
\item the region ($x_3 < x < 1$), where the ratio increases rapidly with increasing $x$. This 
behavior is dominantly a kinematic effect since the free-nucleon cross section vanishes for $x 
\rightarrow 1$. It is partly also due to the Fermi motion of the bound nucleons in the nucleus.
 \end{itemize}

\subsection{Low-$x$ data}
In deep-inelastic scattering from stationary targets, the kinematic region of very low $x$ can 
only be accessed with muon beams, since those can be produced with much higher energies than 
electron beams.
The first of such measurements was performed by EMC-NA28, using a muon detection system at small 
scattering angles down to 2 mrad and nuclear targets of C and Ca~\cite{emc-na38}. This 
experiment demonstrated that shadowing persist also at high values of $Q^2$. The low-$x$ region 
was then explored in detail by NMC with nominal incident muon energies
of 90--200 GeV~\cite{Amaudruz91,Amaudruz92a,Amaudruz92b,Amaudruz92c,Arneodo95,
Arneodo96a,Arneodo96b} and at even lower values of $x$ by E665 at FNAL with
a mean incident muon energy of 470 GeV. 

NMC had the main objective to study the nuclear modification of the structure function ${\rm 
F}_2$ with high precision. Cross section ratios
were measured for nine nuclear species. In one set of measurements \cite
{Amaudruz91,Amaudruz92b,Amaudruz95,Arneodo95} $^4$He, $^6$Li, $^{12}$C and $^{40}$Ca were 
compared to deuterium. All these nuclei are isoscalar and no correction for neutron excess
(Equ. (\ref{gl:f2A})) is necessary. In another set of measurements~\cite{Arneodo96a}, $^9$Be, 
$^{27}$Al,  $^{40}$Ca, $^{56}$Fe, $^{119}$Sn and $^{207}$Pb were compared to carbon. 
One important peculiarity of the NMC experiment was the multiple target arrangement. Targets
of different materials were placed in a row at longitudinally well separated locations along the 
spectrometer axis and exposed simultaneously to the beam. Two such rows, differing in the 
ordering of materials, were placed on a common platform. The rows of targets were positioned in 
the beam in turn by lateral displacement of the platform at approximately 30--60 min intervals. 
With this arrangement beam flux and spectrometer acceptance corrections canceled in the 
determination of cross section ratios.
These high-precision measurements are the low-$x$ counterpart of the E139 data at large $x$.

\begin{figure}[h!] %[p]
\centering
\includegraphics[scale=0.22]{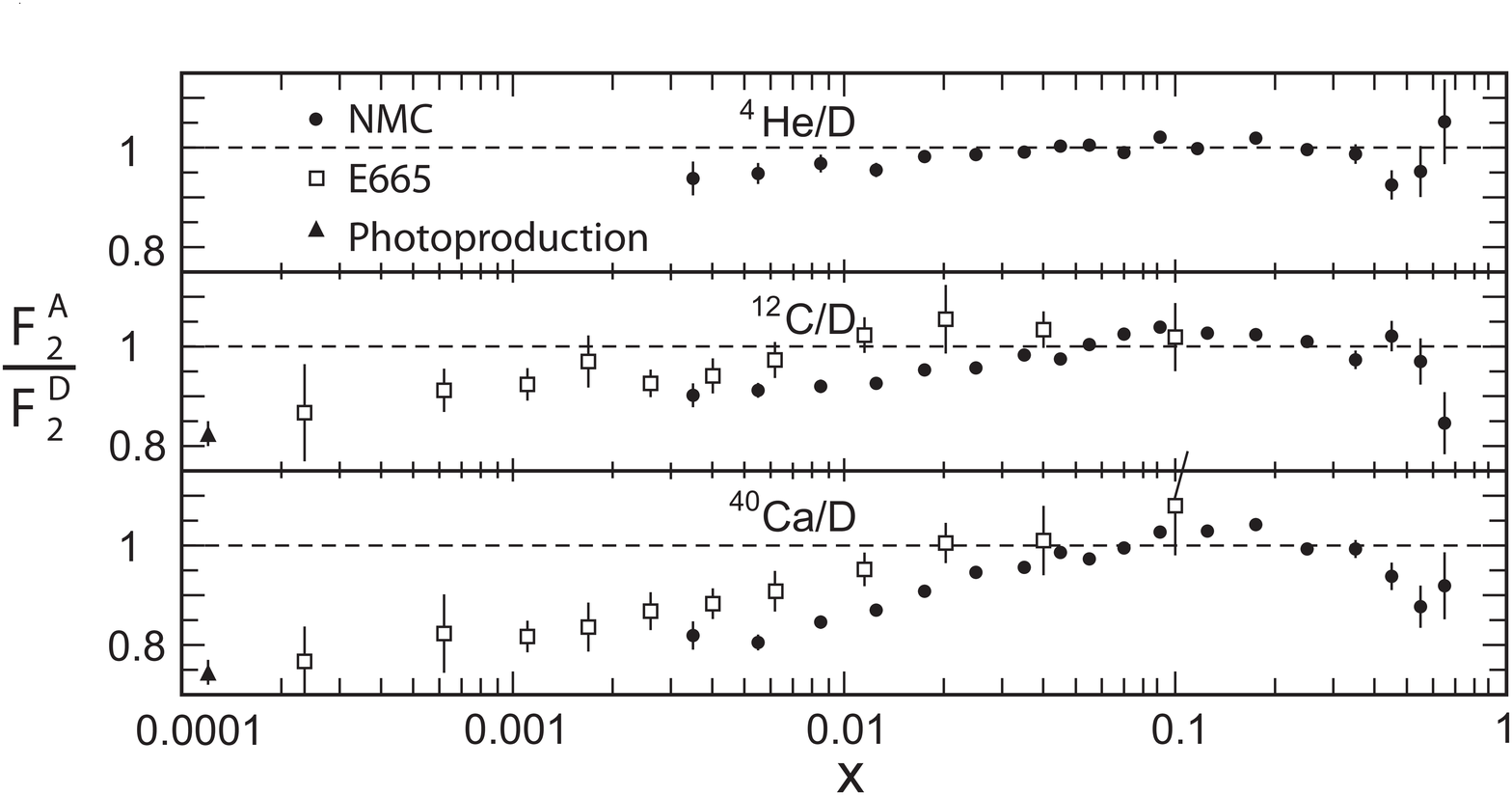}
\caption{%
   The ratio ${\rm F}_2^{\rm A}/{\rm F}_2^{\rm D}$ for He, C and Ca measured at low $x$ by NMC~
\cite{Arneodo95}
and E665~\cite{Adams95}. Note that for both experiments $Q^2 < 1 $ GeV$^2$ for $x$ values below
$\approx 0.005$.
} 
\label{ab:lowxdata}
\end{figure}

As an example the structure function ratios
of $^4$He, $^{12}$C and $^{40}$Ca to deuterium measured by NMC~\cite{Arneodo95} and E665~\cite
{Adams95} are shown in Fig. \ref{ab:lowxdata}. The E665 data nicely extrapolate to the shadowing 
results measured in photoproduction~\cite{Caldwell79}. The NMC and E665 data do not agree very 
well, but the E665 data move downwards by a few percent, when another method of radiative 
corrections is applied~\cite{Adams95}. As already seen in  Fig. \ref{ab:globalx}, the ratio is 
smaller than unity below $x_1 \approx 0.06$ and decreases with decreasing $x$. The effect is 
already visible for $^4$He, it increase with $\rm A$.

\subsection{Nuclear dependence of R}
\label{RAminusRD}
As stated in section \ref{kin}, $\sigma^{\rm A}/\sigma^{\rm D}$ is only equal to 
${\rm F}_2^{\rm A}/{\rm F}_2^{\rm D}$, if
the quantity $\rm R$ 
is independent of the nuclear mass number $\rm A$. 
The  nuclear dependence of $\rm R$ has  been studied by SLAC-E140~\cite{Dasu88,Dasu94}, 
NMC~\cite{Amaudruz92a,Arneodo96b} and HERMES~\cite{Airapetian03}. All measurements are 
consistent with $\rm R$ being independent of $\rm A$. A reanalysis of all SLAC data
with an improved radiative-corrections calculation procedure~\cite{Dasu94} resulted in 
${\rm R}^{\rm Fe} -  {\rm R}^{\rm D} = 0.001 \pm 0.018(stat.) \pm 0.016 (sys.)$,
and the authors conclude that possible contributions to $\rm R$ from nuclear higher-twist 
effects and possible spin-0 constituents in nuclei are not different from those in free nucleons.
This conclusion is supported by the more recent HERMES measurement~\cite{Airapetian03}.   
Averaging over all measurements of $\rm R^{\rm A}/\rm R^{\rm D}$ for light and medium heavy 
nuclei ($^3$He, $^4$He, $^{12}$C, $^{14}$N) HERMES obtained an average value for $\rm R^{\rm 
A}/\rm R^{\rm D}$ of $0.99 \pm 0.03$. The result is unchanged if also the data on the heavier 
nuclei are included in the average.

\subsection{$Q^2$ dependence of the nuclear effects}
Already from the good agreement between the muon and the electron data one can conclude that the 
$Q^2$ dependence of the nuclear effects is very small, since, at the same value of $x$, their 
average $Q^2$  typically differs by more than an order of magnitude.

\begin{figure}[h] %[p]
\centering
\includegraphics[scale=0.20]{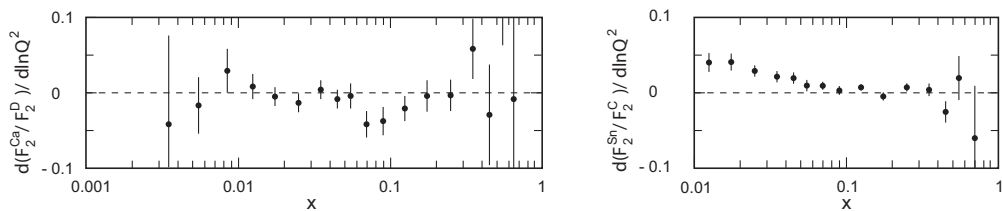}

\caption{%
NMC results for the $x$ dependence of the logarithmic slope 
$d({\rm F}_2^{\rm A}/{\rm F}_2^{\rm D})/d{\rm ln}Q^2$. Left panel: Ca/D~\cite{Arneodo95}, right panel: Sn/C~\cite{Arneodo96b}.
} 
\label{ab:Q2dep}
\end{figure}

The $Q^2$ dependence was studied in some detail by NMC~\cite{Amaudruz95,Arneodo96b}, 
FNAL-E665~\cite{Adams95} and HERMES~\cite{Airapetian03}. As an example, the left panel of 
Fig. \ref{ab:Q2dep} shows the $x$ dependence of the logarithmic slope 
$d({\rm F}_2^{\rm Ca}/{\rm F}_2^{\rm D})/d{\rm ln}Q^2$ from fits of the form 
${\rm F}_2^{\rm A}/{\rm F}_2^{\rm D} = {\rm a} + {\rm b} \cdot {\rm ln}Q^2$ to the NMC data  
\cite{Arneodo95}.  In the covered $Q^2$ range there is little indication for a $Q^2$ dependence 
of the ratio. 
The NMC Sn/C data~\cite{Arneodo96b}, shown in the right panel of  Fig. \ref{ab:Q2dep}, are the 
only exception where one observes an indication of a small $Q^2$ dependence  at small values of 
$x$.

\subsection{Neutrino data}
The available measurements in neutrino/antineutrino-nucleus scattering~\cite{cdhs,neutrinos-bc}
suffer from their large statistical and systematic uncertainties and do not provide additional 
information. Apart from~\cite{cdhs} all these measurements have been performed with bubble 
chambers. Precise measurements of neutrino and antineutrino scattering from deuterium and heavy 
nuclear targets would be very helpful for a separation of nuclear effects in sea-quark and 
valence-quark distributions and a determination of nuclear parton distribution functions 
(nPDFs). But such measurements have to wait for the future realization of a very-high-luminosity 
neutrino factory.

\subsection{Drell-Yan data}

The nuclear modification of anti-quark distributions can also be investigated by the  
Drell-Yan process~\cite{Drell70} in proton-nucleus scattering: ${\rm p} 
{\rm A} \rightarrow ({\it l}+{\it l}^-) {\rm X}$. 
In this process a quark (anti-quark) with the four-momentum fraction $x_1$ from
the beam proton  and an anti-quark (quark) of the target nucleon  with $x_2$ annihilate 
electromagnetically into a
virtual photon, which immediately decays into a charged-lepton pair: ${\rm q}(x_1) 
\bar{\rm q}(x_2) \rightarrow \gamma^* \rightarrow {\it l}^+ {\it l}^-$. 

The longitudinal momentum of the 
${\it l}^+ {\it l}^-$ pair in the proton-nucleon center-of-mass system
is approximately given by
$p_L^{{\it l}^+ {\it l}^-} \approx (x_1 - x_2)\sqrt{s}/2$ and its invariant mass by
$({\rm M}_{{\it l}^+ {\it l}^-})^2 = x_1x_2 s$. Here,  $\sqrt{s}$ is the nucleon-nucleon 
center-of-mass 
energy.
The Drell-Yan cross section reads
\begin{equation}
\frac {d^2\sigma}{d x_1 d x_2} 
= {\rm K} \frac{4\pi\alpha^2 }{9x_1x_2} \frac{1}{s}
 \sum_{\rm q=u,d,s...} z_{\rm q}^2 \left[ {\rm q}(x_1)\bar{\rm q}(x_2) + 
\bar{\rm q}(x_1){\rm q}(x_2) \right] \;,
\label{gl:DrellYan}
\end{equation}
where ${\rm K} \approx 2$ is a factor representing the deviation from the simple parton model 
due to QCD corrections. By a suitable choice of the kinematics of the 
$\ell^+ \ell^-$ pair the  
quark and anti-quark distributions in the target can be determined separately. If one requires 
for instance $x_1-x_2 > 0.3$, then the second term in (~\ref{gl:DrellYan}) can be neglected and 
the cross section ratio for two nuclei $A1$ and $A2$ is to a good approximation equal to the
ratio of anti-quark distributions $\bar{\rm q}^{A1}/\bar{\rm q}^{A2}$ in the target.

\begin{figure}[h!] %[p]
\centering
\includegraphics[scale=0.18]{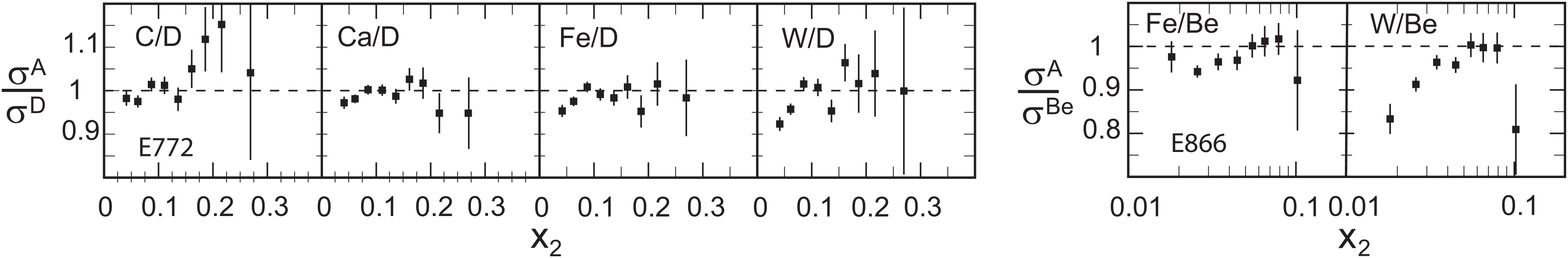}
\caption{%
Drell-Yan results for the cross-section ratio $\sigma^{\rm A}/\sigma^{\rm D}$ \cite{Alde90} 
(left) and 
$\sigma^{\rm A}/\sigma^{\rm Be}$ \cite{Vasiliev99} (right)as a function of $x_2$.
} 
\label{ab:DY}
\end{figure}

In the left panel of Fig. \ref{ab:DY}, $\sigma^{\rm A}/\sigma^{\rm D}$ for four nuclei
(C, Ca, Fe, W) measured by FNAL-E772~\cite{Alde90} is shown as a function of $x_2$, whereas the 
right panel shows $\sigma^{\rm A}/\sigma^{\rm Be}$ for (Fe, W) measured by FNAL-E866~\cite
{Vasiliev99}, both with a proton-beam energy of 800 GeV. At the smallest $x_2$ values 
the data show an indication of shadowing, while above $x_2 \approx 0.06$ the ratios are 
consistent with unity (with the exception of the C/D data) leading to the conclusion that 
anti-shadowing is very likely not a sea-quark effect or caused by nuclear pions. This aspect 
will be studied in detail by the experiment E906/Seaquest at the 120 GeV beam of the FNAL Main 
Injector \cite{Reimer11}. 

Unfortunately the  Drell-Yan cross section is very small and it is very difficult to study the 
process with colliding ion beams at much higher center-of-mass energies. Nevertheless such 
measurements in proton-deuteron and proton-nucleus collisions at RHIC or the LHC would be very 
desirable for the determination of nPDFs at low values of $x$.

\subsection{Dependence on nuclear properties}
\label{nuclprop}

\noindent {\bf Dependence on nuclear mass ${\rm A}$.}
From Figs. \ref{ab:SLACE139} and \ref{ab:lowxdata} it is obvious that the nuclear effects 
increase continuously with nuclear mass number ${\rm A}$. This aspect has been studied in detail 
by E139 \cite{Gomez94} and NMC \cite{Arneodo96a}. 

\begin{figure}[h!] %[p]
\centering
\includegraphics[scale=0.18]{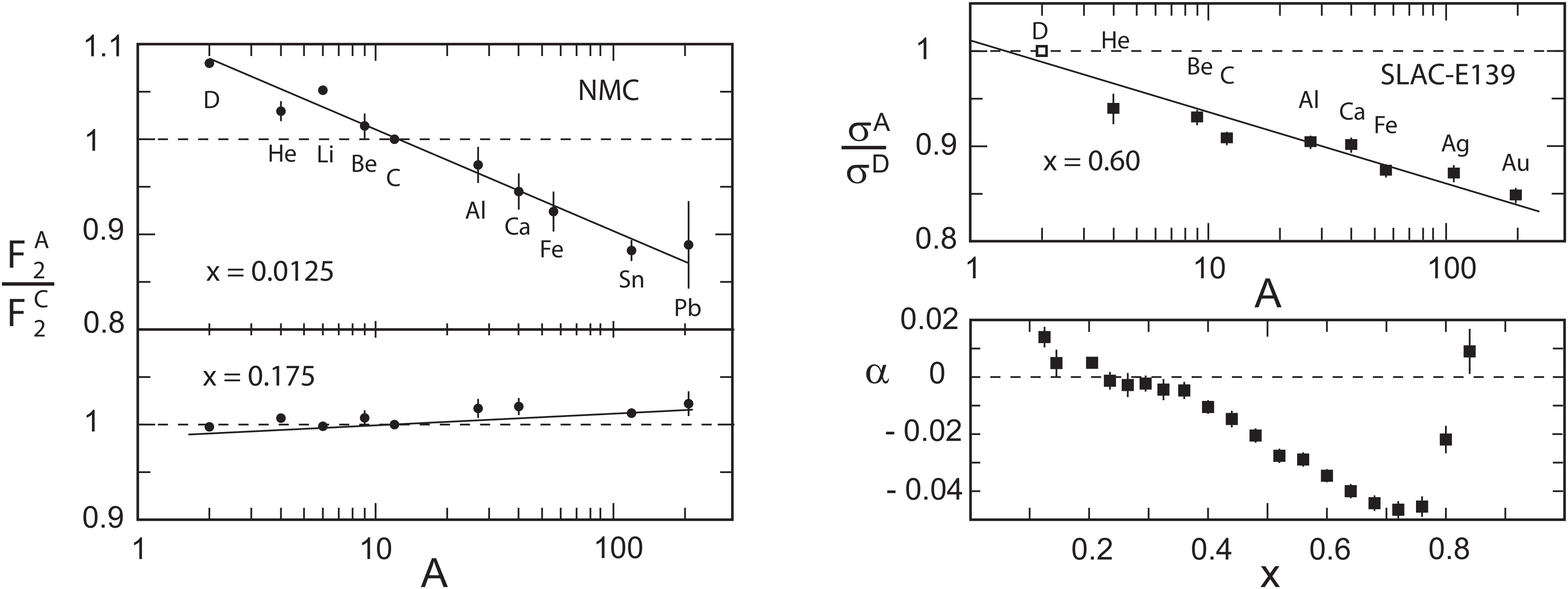}
\caption{%
${\rm F}_2^{\rm A}/{\rm F}_2^{\rm C}$ as a function of nuclear mass $A$ at low $x$ from NMC \cite
{Arneodo96a} (left) and $\sigma^{\rm A}/\sigma^{\rm D}$ at high $x$ from SLAC-E139 \cite
{Gomez94}(right). The lower right panel shows the coefficient
$\alpha(x)$ from a fit of the form $\sigma^A/\sigma^D(x) = c(x)A^{\alpha(x)}$.
} 
\label{ab:Adep}
\end{figure}

The left panel of Fig. \ref{ab:Adep} shows the NMC results \cite{Arneodo96a} for the dependence 
of ${\rm F}_2^{\rm A}/{\rm F}_2^{\rm C}$ on ${\rm A}$ for the two bins $x = 
0.0125$ and $x = 0.175$, and the upper plot in the right panel shows the E139 results \cite
{Gomez94} for $\sigma^{\rm A}/\sigma^{\rm D}$ in the bin $x = 0.60$. Obviously 
the nuclear effects increase to a good approximation linearly with log${\rm A}$. Small 
deviations from this linear dependence are observed in the left panel for He and Li and in the 
right panel for He and C. Obviously there are other nuclear properties than A affecting the 
nuclear dependence. The lines are results of fits of the form:  
$ {\rm Ratio}= c(x){\rm A}^{\alpha(x)}$. 
The coefficient ${\alpha(x)}$ determined from the E139
data is shown in the lower right panel as a function of $x$. If we rewrite $\alpha(x)$ as
$\alpha(x) = \alpha'(x) - 1$, then ${\rm A}^{\alpha'(x)}$ can be interpreted as the effective 
number of nucleons participating in the interaction. 

\bigskip

\begin{figure}[h!] %[p]
\centering
\includegraphics[scale=0.18]{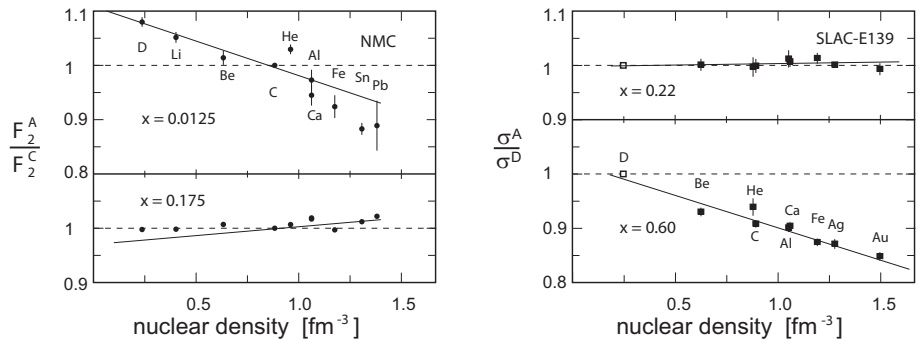}
\caption{%
${\rm F}_2^{\rm A}/{\rm F}_2^{\rm C}$ as a function of nuclear density $\rho$ at low $x$ from 
NMC \cite{Arneodo96a} and $\sigma^{\rm A}/\sigma^{\rm D}$ at high $x$ from SLAC-E139 \cite
{Gomez94} .
} 
\label{ab:rhodep}
\end{figure}

\noindent {\bf Dependence on nuclear density $\rho$.}
One important aspect for the understanding of the nuclear medium effects is their dependence on 
the nuclear density $\rho$. In Fig. \ref{ab:rhodep} the data presented in Fig. \ref{ab:Adep}
are shown as a function of $\rho$ in four bins of $x$.

Here $\rho({\rm A})$ is given by $\rho({\rm A}) = 3{\rm A}/4\pi {\rm R}_e^3$, with
${\rm R}_e^2 = 5<r^2>/3$, and $<r^2>$ is taken from Ref. \cite{Vries87}. At large $x$ the
cross section ratio approximately scales with $\rho$, with the exception of
the rather special nuclei $^4$He and $^9$Be. The deviation of $^9$Be from this linear behavior 
is, however, much smaller than in the analysis of JLAB-E03103 \cite{Seely09}, where a density
based on ab initio few-body calculations \cite{Pieper01}, scaled by $({\rm A}-1)/{\rm A}$ is used.
The small-$x$ data are not well described by a linear function of nuclear density.

\bigskip

\noindent {\bf Dependence on nuclear radius $r$.}
The deviations from a linear behavior observed in Fig. \ref{ab:rhodep} indicate that besides the 
nuclear density other parameters like the nuclear radius or the nuclear surface may play a role. 
The nuclei $^4$He, $^6$Li, $^{12}$C and $^{40}$Ca have been used by NMC 
\cite{Amaudruz91,Amaudruz92a,Amaudruz95,Arneodo95}  to possibly differentiate between effects 
originating from the nuclear density or from the nuclear radius.
These nuclei differ primarily either in radius ($r$) or density ($\rho$). In particular,  
$^{6}$Li ($r = 2.6 ~{\rm fm}, ~\rho = 0.04 ~{\rm fm}^{-3}$)
and $^{12}$C ($r = 2.5 ~{\rm fm}, ~\rho = 0.09 ~{\rm fm}^{-3}$) have nearly equal radii but 
different densities, whereas $^4$He ($r = 1.7 ~{\rm fm}, ~\rho = 0.09 ~{\rm fm}^{-3}$) and
$^{12}$C have nearly equal densities but different radii.  $^{12}$C and $^{40}$Ca ($r = 3.5 
~{\rm fm}, ~\rho = 0.11 ~{\rm fm}^{-3}$)
differ more in radius than in density. The analysis presented in \cite{Amaudruz92a} demonstrates 
that there is a rather complicated interplay between the dependences on radius and density:
the depletion at low $x$ is larger in $^{12}$C than in $^{6}$Li, showing that at the same radius 
the effect increases with $\rho$; the comparison of the pairs ($^4$He,$^{12}$C)  and 
($^{40}$Ca,$^{12}$C) indicates that at similar densities the effect increases with radius; the 
depletion in $^{40}$Ca is twice as large than in $^{6}$Li, implying that the depletion increases 
with both radius and density; and the Li/He ratio is
consistent with unity over the common $x$ range, indicating that the opposing dependencies
on radius and density tend to cancel. Furthermore NMC has shown that at low $x$ the data are 
best described by a fit of the form 
${\rm F}_2^{\rm A}/{\rm F}_2^{\rm C} = a + b{\rm A}^{-1/3} + c {\rm A}^{-2/3}$~\cite{Arneodo96a}. 

\section{Interpretations}
It is beyond the scope of this lecture note to address the multitude of possible explanations for
the observed nuclear medium effects (for a rather detailed discussion see Ref. \cite{Norton03}). 
Instead, only the main ideas of some classes of models will be summarized without a discussion 
why and where they fail to reproduce the data correctly. 

\subsection{Shadowing}
\label{shad}
The first class of models deals with the shadowing region.
The term `shadowing' has been introduced to explain the reduction of the nuclear cross sections in
photoproduction, but has then also been used in the discussion of the low-$x$ modification of 
the nuclear cross section in inelastic lepton-nucleon scattering. There are two possibilities to 
explain this phenomenon.
In the first approach the parton distributions in bound nucleons remain unchanged compared to 
those in the free nucleon. The interaction is viewed in the rest system of the nucleus. The 
nuclear effects are attributed to a modification of the interaction of the virtual photon with 
the atomic nucleus by fluctuations of the virtual photon into quark-antiquark pairs. 
Such a pair then interacts with the nucleus via the {\it strong} interaction. Since the strength 
of the latter is much larger than the electromagnetic one, the interaction does no longer happen 
incoherently with all the nucleons in the nucleus but preferentially with those at the front 
surface. The nucleons being in the `shadow' of the nucleons at the front surface then do not or 
much less contribute to the cross section. Quantitatively this happens, when the fluctuation 
length 
$\Delta{d} \approx 1/Mx \approx 0.2~{\rm fm}/x$ is larger than the mean free path ${\rm L} \cong 
2.5 ~{\rm fm}$ of the
quark-antiquark pair, i.e., for $x$ below 
$\approx 0.08$.

In the second approach the effect is attributed to a modification of the quark and gluon 
distributions in the nucleus. The interaction of the virtual photon with the nucleus is viewed 
in a fast moving system, where the nucleon with diameter $\rm D$ is  Lorentz contracted to a 
disc of thickness ${\rm D}' \approx {\rm D} \cdot {\rm M}/|\vec{{\rm P}}|$ and the mean nucleon 
distance ${\rm d} \approx 2 ~{\rm fm}$ to
${\rm d}' = {\rm d} \cdot M/|\vec{P}|$. 
The longitudinal position of its constituents, however, has an uncertainty of 
$\Delta{z} = 1/x|\vec{P}|$. 
At small values of $x$ this uncertainty $\Delta{z}$ can be much larger than 
$\rm D'$  
and in a nucleus much larger than $\rm d'$.
For $x < 1/d \cdot 1/{\rm M} \approx 0.1$ there will be a spatial overlap of sea quarks and 
gluons of different nucleons.  
The smaller $x$, the larger is the number of nucleons sharing their contents of
sea-quarks and gluons. The effect increases with increasing mass number ${\rm A}$ and saturates 
for 
$x \approx 1/{\rm D}_{\rm A} \cdot 1/{\rm M}$, where ${\rm D}_{\rm A}$ is the diameter
of the nucleus in its rest frame.  Thus, the density of gluons and sea quarks at the position of 
a nucleon in a nucleus can be much larger than for a free nucleon.
Due to this  `overcrowding'  the probability for an interaction between sea quarks and gluons is 
increased and by pair annihilation their density is reduced again,
 resulting in the observed reduction of the nuclear structure function. 
Momentum conservation requires that this reduction of the number of partons at low values of $x$ 
gets compensated by an enhancement at larger $x$, i. e., anti-shadowing.

Both approaches for shadowing are equivalent. They describe the same phenomenon but viewed in a 
different reference frame. More details  can be found, e. g., in Refs. \cite
{Kopeliovich13,Frankfurt12a}.

\subsection{Convolution Models}
In the second class of models the structure function of a nucleus $\rm A$ is described as 
the incoherent sum over contributions of all kind of clusters $\rm c$ with structure functions 
${\rm F}_2^{\rm c}(x/y)$ convoluted with the probability $f_{\rm c}^{\rm A}(y)$ to find a certain 
cluster $\rm c$ of momentum $y$ in the nucleus: 

\begin{equation}
F_2^{\rm A}(x,Q^2) = \sum_{\rm c} \int_x^A dy f_{\rm c}^{\rm A}(y) F_2^{\rm c}(x/y) .
\end{equation}

\noindent Examples for such clusters are the nucleon itself, undisturbed or with a reduced mass 
due to 
nuclear binding or with an increased size due to different boundary conditions in the 
nuclear 
environment, extra pions being responsible for nuclear binding, $\Delta$-isobars, multi-quark
clusters like bags of 6 quarks, 9 quarks or 12 quarks (i.e., $\alpha$ particles) or the whole 
nucleus as one big bag 
with free quark and color flow throughout the whole nuclear volume.

There is a lot of freedom in these approaches, concerning as well the choice of 
$f_{\rm c}^{\rm A}(y)$ as the parameterization of $F_2^{\rm c}(x/y)$ that both are badly known. 
Consequently, it is not very surprising
that they succeed to reproduce a portion of the data rather well, at least in the medium $x$ 
range.

\subsection{Rescaling models}
In the third class of approaches the EMC effect is explained by a change of either the $Q^2$ 
scale or the $x$ scale for the nuclear structure function compared to the free nucleon's one.

\medskip

\noindent{\bf $Q^2$ rescaling.}
In $Q^2$ rescaling models, first proposed in Refs. ~\cite{Close83} and 
\cite{Nachtmann84}, the EMC effect is related to a change of confinement size inside the nucleus. 
The
{\it A dependence} of quark and gluon distributions for bound nucleons and the {\it $Q^2$ 
evolution} of these distributions for free nucleons both have the same origin. They are caused 
by the color forces between quarks and gluons which ensure confinement and are the origin of 
scaling violations, i.e., the increase of the structure function with $Q^2$ at low values of $x$ 
and the decrease with increasing $Q^2$ at large values of $x$. The qualitative argument is as 
follows: The strength of the strong force between quarks is not only determined by the 
transverse resolution
$1/\sqrt{Q^2}$ at which they are probed, but also by the radial extension $r_{\rm A}$ of the 
volume in which they are confined. Therefore, the relevant parameter for the strength of the 
strong coupling constant $\alpha_{\rm s}$ is not just $Q^2$ but $(Q \cdot r_{\rm A})^2$. (This 
is similar to the situation in nuclear physics where the form factors for spherical nuclei have 
the identical
oscillating pattern when plotted against $(Q \cdot r_{\rm A})$). If the confinement size is 
modified inside the nucleus, either due to a `swelling' of nucleons, the formation of multi-quark 
bags, short-range nucleon-nucleon correlations or free quark and color flow throughout the whole 
nucleus, then, as a consequence, quark and gluon distributions obtained for nuclei $A$ and $B$ 
are related by 
\begin{equation}
q_{\rm A}(x,Q^2) = q_{\rm B}(x,\xi \cdot Q^2),\;  g_{\rm A}(x,Q^2) = g_{\rm B}(x,\xi \cdot Q^2).
\end{equation}
$\xi$ is a rescaling parameter determined by the two confinement scales 
$r_{\rm A}$ and $r_{\rm B}$. In the so-called 'dynamical rescaling' models
it is given by
\begin{equation}
\xi = (r_{\rm A}/r_{\rm B})^{2\alpha_s(\mu_A^2)/\alpha_s(Q^2)} ,
\end{equation}
where $\mu_A$ is a low-momentum cut-off for radiating gluons.
Consequently, at the same value of $Q^2$, $F_2^A/F_2^B > 1$ for small values of $x$ and 
$F_2^A/F_2^B < 1$ for large values of $x$, if $r_A > r_B$.
\medskip

\noindent{\bf $x$ rescaling.}
In the $x$-rescaling models, first proposed in \cite{Garcia84,Staszel84} and then refined by 
numerous authors, the depletion of the nuclear structure function at medium $x$ is explained by 
conventional nuclear binding and Fermi-motion corrections. The $x$ dependence can be 
reasonably well reproduced if, for a nucleus, the scaling variable $x$ is replaced by a modified 
one $x^* > x$.
For a nucleon $i$ moving with momentum 
$\vec{\rm p}_i$ in a nucleus the variable $x = Q^2/2{\rm M}\nu$ has to be replaced by
\begin{equation}
x_i = Q^2/2{\rm p}_i q = Q^2/\left [2({\rm M}+{\rm E}_i)\nu - 2 \vec{\rm p}_i\vec{q} \right ],
\end{equation}
where ${\rm E}_i$ is the removal energy of the nucleon ($<{\rm E}_i> \cong -25$ MeV) and 
$\vec{q}$ is the momentum of the virtual photon. 
Consequently, at the same kinematics of the scattered lepton, the effective $x^*$ at which the 
structure function is probed in a nucleus, is larger than $x$ for a free nucleon. At large $x$ 
the structure function ${\rm F}_2$ is steeply falling with $x$ and a depletion of the 
structure function ratios is naturally explained.

\section{The new ingredient: $x > 1$}
\label{xlargerone}
Recently their has been renewed interest in the EMC effect and its possible origin by JLAB 
measurements of the cross section ratios in inclusive electron scattering ${\rm A}(e,e')$
in the kinematic region $x > 1$ where the cross section vanishes for scattering from free 
nucleons. 

\begin{figure}[h!] %[p]
\centering
\includegraphics[scale=0.18]{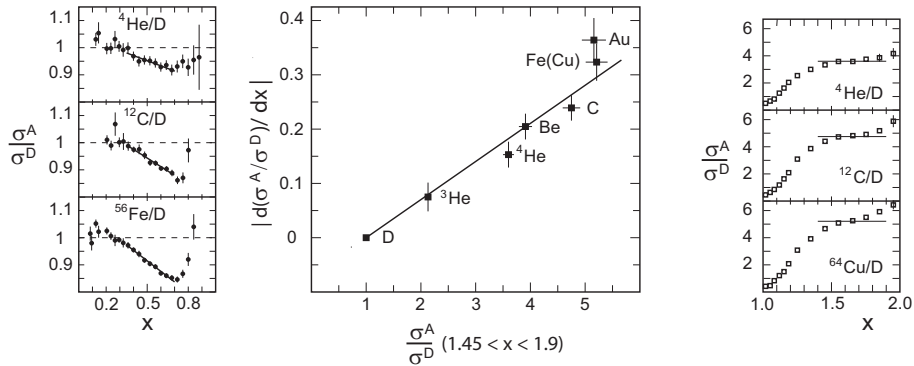}
\caption{%
The cross section ratio $\sigma^{\rm A}/\sigma^{\rm D}$ measured by SLAC-E139 at
$x < 1$ (left panel) and by JLAB-E02019 in the region $1 <x < 2$ (right panel). The middle panel 
shows the correlation between the slope $|d(\sigma^{A}/\sigma^{D})/dx|$ in the region $0.35 
\leq  x \leq  0.7$
  and the height of the plateau of  $\sigma_A/\sigma_{D}$ in the region $1.45 < x < 1.9$ (after 
\cite{Weinstein11}). 
} 
\label{ab:SRC}
\end{figure}

The CLAS experiment~\cite{Egyan03,Egyan06} has measured $\sigma^{\rm A}/\sigma^{^3{\rm He}}$
for $^4$He,  $^{12}$C and $^{56}$Fe and JLAB-E02019 ~\cite{Fomin12} measured 
$\sigma^{\rm A}/\sigma^{\rm D}$ for $^3$He, $^4$He, Be, C, Cu and Au. In the right panel
of Fig. \ref{ab:SRC} the cross section ratios $^4$He/D, C/D and Cu/D are shown as a function of 
$x$. They rise with $x$ until they reach a plateau at $x \approx 1.4-1.5$. The height of this 
plateau increases with ${\rm A}$. Such a behavior has already been observed with less accuracy 
by an experiment at SLAC ~\cite{Frankfurt93}.
A similar pattern is seen in the CLAS comparison of the nuclear cross sections to $^3$He. Here 
the measurements extend up to $x = 3$ and an indication of a second plateau is observed for $x > 
2.25$. Such a behavior is being interpreted as the manifestation
of short-range nucleon-nucleon (mostly p--n) correlations  (three-nucleon correlations for $x$ 
beyond 2) or, in another approach, as the ratio of the probabilities to find 6-quark or 9-quark 
clusters in nuclei compared to the reference nucleus.

Fig. \ref{ab:SRC} shows a very interesting observation \cite{Weinstein11}. In the left panel the 
E139 cross section ratios $^4$He/D, C/D and Fe/D
are shown for $x < 1$. The lines correspond to the slopes $d(\sigma^{A}/\sigma^{D})/dx$ in the 
region $0.35 \leq  x \leq  0.7$, that characterize the strength of the EMC effect in this region 
and are unaffected by overall normalization uncertainties. In the middle panel the slopes
of the E139 data (tabulated in \cite{Weinstein11} and corrected by 
$x_{\rm A} = x_{\rm p} \cdot {\rm A}{\rm M}_{\rm p}/{\rm M}_{\rm A}$ \cite{Frankfurt12b}) are 
plotted against the height of the plateaus~\cite{Fomin12}. 
Obviously there is a strong correlation between these quantities as indicated by the straight 
line. It is rather unlikely that this correlation is purely accidental and one can therefore 
rather safely assume that a large fraction of the strength of the EMC effect in the valence 
quark region is due to short-range nucleon-nucleon correlations.

\section{Summary and Outlook}
The EMC effect is with us now for 30 years. It has stimulated huge experimental and theoretical 
efforts, but its origin is still not fully understood. Recent data shed some new light on its 
possible origin, i.e., short-range nucleon-nucleon correlations may play an important role for 
the observed nuclear modifications. Still more precise data are needed, especially on the 
nuclear gluon and antiquark distributions at very low $x$ to constrain the initial state for the
${\rm AA}$ program at RHIC and LHC. These hopefully will come from future measurements at JLAB12,
RHIC and LHC and eventually also from the proposed projects EIC and LHeC.

\end{document}